\documentclass[12pt,a4paper]{article}
\usepackage{epsfig}
\usepackage{mathrsfs}
\usepackage{epsfig}
\begin{document}
\title{Lepton flavor violating signals of the LHT model via $e^{+}e^{-}$ and $\gamma\gamma$ collisions at the ILC}
\author{Wei Ma, Chong-Xing Yue, Jiao Zhang, and Yan-Bin Sun\\
{\small Department of Physics, Liaoning  Normal University, Dalian,
116029 People's Republic of China}
\thanks{cxyue@lnnu.edu.cn}}
\date{\today}
\maketitle

\begin{abstract}

Taking into account the constraints on the free parameters of the
littlest Higgs model with $T$ parity (called the LHT model) from
some rare decay processes, such as $\mu\rightarrow e\gamma$ and
$\mu\rightarrow 3e$, we consider the contributions of the LHT model
to the lepton flavor violating (LFV) processes
$e^{+}e^{-}\rightarrow l_{i}\bar{l}_{j}$ and
$\gamma\gamma\rightarrow l_{i}\bar{l}_{j}$ ($i\neq j$). We find that
the LHT model can indeed produce significant contributions to these
LFV processes and its LFV signal might have a chance of being
observed in the future International Linear Collider experiments.

\vspace{1cm}

PACS number:  12.60.Cn, 11.30.Fs, 13.66.De

\end {abstract}

\vspace{0.8cm}
\newpage

\section*{I. Introduction}
\hspace{0.5cm}During the past decade, neutrino oscillation
experiments have provided us with very convincing evidence that
neutrinos are massive particles mixing with each other [1].
Moreover, their masses are extremely small, while mixing is nearly
maximal, which means that the lepton flavor violating (LFV)
processes are allowed. However, it is well known that, in the
standard model (SM), neutrinos are massless and the LFV processes
are not allowed at tree level. Thus, the exciting experimental fact
opens a window to new physics beyond the SM [2]. In fact, lepton
flavor symmetry is an accidental symmetry at low energy, and it may
be violated beyond the SM. Many kinds of popular specific models,
like supersymmetry, technicolor, and little Higgs models, indicate
the possibility of large LFV. Therefore, an LFV signal in the
charged lepton sector would be a clear hint for new physics beyond
the SM. Experimental detection of the LFV phenomenon can provide an
evidence of new physics.

Searching for new physics beyond the SM is one of the most important
issues of current particle physics. The CERN Large Hadron Collider
(LHC) can generate very massive new particles and will essentially
enlarge the possibilities of testing for new physics effects.
However, the LHC is a hadronic machine and precision measurements
will be quite hard to undertake there. Also, the existence of large
backgrounds at the LHC may hinder discoveries of new physical
phenomena already possible at the energies that this accelerator
will achieve. Thus, the next generation $e^{+}e^{-}$ International
Linear Collider (ILC) with the center of mass (c.m.) energy
$\sqrt{s}=0.5 - 1TeV$ and the typical integrated luminosity
$\pounds_{int}=0.5-1$ $ab^{-1}$ is currently being designed [3, 4].
Because of its rather clean environment and high luminosity, the ILC
will allow unambiguous precision measurements. In such a collider,
in addition to $e^{+}e^{-}$ collision, one can also realize
$\gamma\gamma$ collision with the photon beams generated by the
backward Compton scattering of incident electron- and laser- beams.
The $\gamma\gamma$ collision offers a unique opportunity to explore
new physics effects through production mechanisms which are not
accessible in leptonic or hadronic machines [5].

It is well known that many popular models beyond the SM predict the
presence of new particles, such as new gauge bosons, new fermions,
and new scalars, which can generally enhance the branching ratios
for some LFV decay processes, for instance $l_{i}\rightarrow
l_{j}\gamma$, $l_{i}\rightarrow l_{j}l_{k}\bar{l}_{l}$, and
$Z\rightarrow l_{i}\bar{l}_{j}$. The upper experimental limits for
some of these LFV decay processes can give serve constraints on the
corresponding new physics models. Nevertheless, it is possible that
the LFV signals may be observed at the ILC. This fact has lead to
lot of work to study the LFV processes $e^{+}e^{-}\rightarrow
l_{i}\bar{l}_{j}$ and $\gamma\gamma\rightarrow l_{i}\bar{l}_{j}$ in
the framework of specific models beyond the SM [6, 7] and in a
model-independent manner [8]. Taking into account the constraints of
the upper experimental limits for some LFV decay processes on the
littlest Higgs model with $T$ parity (called the LHT model) [9], in
this paper, we will focus our attention on its contributions to the
LFV processes $e^{+}e^{-}\rightarrow l_{i}\bar{l}_{j}$ and
$\gamma\gamma\rightarrow l_{i}\bar{l}_{j}$ ($i\neq j$). We calculate
the production cross sections of these processes induced by the LHT
model and discuss the possibility of detecting the LFV signals of
the LHT model via $e^{+}e^{-}$ and $\gamma\gamma$ collisions at the
ILC.

The rest of this paper is organized as follows. In Sec. II, we
briefly review the essential features of the LHT model, which are
related our calculation. A simple discussion about the constraints
on the LHT model from some LFV decay processes is also given in this
section. The production cross sections of the LFV processes
$e^{+}e^{-}\rightarrow l_{i}\bar{l}_{j}$ and
$\gamma\gamma\rightarrow l_{i}\bar{l}_{j}$ are calculated in Secs.
III and IV, respectively. Finally, the conclusions are given in Sec.
V.

\section*{II. The essential features of the LHT model}

\hspace{0.5cm}Little Higgs theory [10] was proposed as an
alternative solution to the hierarchy problem of the SM, which
provides a possible kind of  electroweak symmetry breaking (EWSB)
mechanism accomplished by a naturally light Higgs boson. In order to
make the littlest Higgs model consistent with electroweak precision
tests and simultaneously having the new particles of this model in
the reach of the LHC, a discrete symmetry, T-parity, has been
introduced, which forms the LHT model. The detailed description of
the LHT model can be found for instance in Refs.[9,11,12], and here
we just want to briefly review its essential features, which are
related to our calculation.

The LHT model is based on an $SU(5)/SO(5)$ global symmetry breaking
pattern. A subgroup $[SU(2)\times U(1)]_{1}\times [SU(2)\times
U(1)]_{2}$ of the $SU(5)$ global symmetry is gauged, and at the
scale $f$ it is broken into the SM electroweak symmetry
$SU(2)_{L}\times U(1)_{Y}$. T-parity exchanges the $[SU(2)\times
U(1)]_{1}$ and $[SU(2)\times U(1)]_{2}$ gauge symmetries. The T-even
combinations of the gauge fields are the SM electroweak gauge bosons
$W^{a}_{\mu}$ and $A_{\mu}$. The $T$-odd combinations are T-parity
partners of the SM electroweak gauge bosons.

After taking into account EWSB, at the order of $v^{2}/f^{2}$, the
masses of the $T$-odd set of the $SU(2)\times U(1)$ gauge bosons are
given  as
\begin{eqnarray}
M_{B_{H}}=\frac{g_{1}f}{\sqrt{5}}[1-\frac{5v^{2}}{f^{2}}],\hspace{0.5cm}M_{Z_{H}}\approx
M_{W_{H}}=g_{2}f[1-\frac{v^{2}}{8f^{2}}],
\end{eqnarray}
where $v=246GeV$ is the electroweak scale and $f$ is the scale
parameter of the gauge symmetry breaking of the LHT model, $g_{1}$
and $g_{2}$ are the SM $U(1)_{Y}$ and $SU(2)_{L}$ gauge coupling
constants, respectively.

A consistent implementation of T-parity also requires the
introduction of mirror fermions --- one for each quark and lepton
species. The masses of the $T$-odd (mirror) fermions can be written
in a unified manner:
\begin{eqnarray}
M_{F_{i}}=\sqrt{2}k_{i}f,
\end{eqnarray}
where $k_{i}$ are the eigenvalues of the mass matrix $k$ and their
values are generally dependent on the fermion species $i$. These new
fermions ($T$-odd quarks and $T$-odd leptons) have new flavor
violating
 interactions with the SM fermions mediated by the new gauge
 bosons ($B_{H}, W_{H}^{\pm}$ and $Z_{H}$) and at higher order by the
 triplet scalar $\Phi$. These interactions are governed by new
 mixing matrices $V_{Hd}$ and $V_{Hl}$ for down-type quarks and
 charged leptons, respectively. The corresponding matrices in the
 up-type quarks
 ($V_{Hu}$) and neutrino ($V_{H\nu}$) sectors are obtained by means of
 the relations:
\begin{eqnarray}
V_{Hu}^{+}V_{Hd}=V_{CKM}, \hspace{0.5cm}V_{H\nu}^{+}V_{Hl}=V_{PMNS}.
\end{eqnarray}
Where the Cabibbo-Kobayashi-Maskawa (CKM) matrix $V_{CKM}$ is
defined through flavor mixing in the down-type quark sector, while
the PMNS matrix $V_{PMNS}$ is defined through neutrino mixing.

The Feynman rules of the LHT model have been studied in Ref.[12] and
the corrected Feynman rules of Ref.[12] are given in Refs.[13, 14].
To simplify our paper, we do not list them here.

From the above discussions, we can see that the flavor structure of
the LHT model is much richer than the one of the SM, mainly due to
the presence of three doublets of mirror quarks and leptons and
their interactions with the ordinary quarks and leptons. Such new
flavor changing interactions can induce that the LHT model generates
contributions to some flavor changing processes. The contributions
of the LHT model to the LFV decay processes have been extensively
studied and compared with current experimental limits in the
literature [14,15,16]. It has been shown that the LHT model can
enhance the SM prediction values by several orders of magnitude and
the experimental measurement data for some LFV decay processes can
give serve constraints on the free parameters of the LHT model.

The new particles predicted by the LHT model can generate
significative contributions to the LFV decay process
$\mu\rightarrow3e$ via box diagrams and the effective vertices
$\gamma(Z)e\bar{e}$ generated by $\gamma-$ and $Z-$penguins, while
only via $\gamma-$ penguin for the LFV decay process $\mu\rightarrow
e\gamma$ [14, 15]. In order to suppress the values of the branching
ratios $Br(\mu\rightarrow e\gamma)$ and $Br(\mu\rightarrow3e)$
predicted by the LHT model below the present experimental upper
bounds, the relevant mixing matrix $V_{Hl}$ must be rather
hierarchical or mass splitting for the first and second $T$-odd
lepton masses is very small. Reference [15] has shown that, for
$f=1TeV$, the mass splitting $\triangle M = \mid M_{l_{H}^{1}}-
M_{l_{H}^{2}}\mid$ must satisfy $\triangle M\leq 40GeV $ and $
60GeV$ for assuming $V_{Hl} = V_{PMNS}$ and $V_{Hl} = V_{CKM}$,
respectively, in which $M_{l_{H}^{i}}$ is the mass of the $i$-th
generation $T$-odd lepton. It is should be noted that the value of
$\triangle M $ increases as the scale parameter $f$ increases.
However, the present experimental upper bounds for
$\mu\rightarrow3e$ and $\mu\rightarrow e\gamma$ can not give
significative constraints on the mass of the third $T$-odd lepton
generation. Thus, in this paper, we will assume
$M_{l_{H}^{1}}=M_{l_{H}^{2}}=M_{2}$ and $M_{l_{H}^{3}}=M_{3}$. In
our following numerical estimation, we take the mass parameters
$M_{2}$ and $M_{3}$, and the scale parameter $f$ as free parameters.

\section*{III. The LFV processes $e^{+}e^{-}\rightarrow l_{i}\bar{l}_{j}$ in the LHT model}

\hspace{0.5cm}From above discussions we can see that the LHT model
can contribute to the LFV processes $e^{+}e^{-}\rightarrow
l_{i}\bar{l}_{j}$ ($i\neq j$) via the effective vertices
$\gamma(Z)l_{i}\bar{l}_{j}$ and the box diagrams. The Feynman
diagrams for the LFV processes $e^{+}e^{-}\rightarrow
l_{i}\bar{l}_{j}$ are depicted in Fig.1, in which $\omega^{\pm,0}$
and $\eta$ represent the would be Goldstone bosons arising from the
LHT model. It is obvious that the contributions of the LHT model to
this process mainly come from the effective vertices
$\gamma(Z)l_{i}\bar{l}_{j}$ [Fig.1(a)--Fig.1(k)], which are related
the LFV processes $\mu\rightarrow3e$ and $\mu\rightarrow e\gamma$.
Thus, the production cross sections of the
 processes $e^{+}e^{-}\rightarrow l_{i}\bar{l}_{j}$ should also be
 constrained by the present experimental upper bounds of these processes.
 However, as shown in Refs.[14, 15], as long as $\triangle M \approx 0$, the LHT model
 can satisfy the constrains from $\mu\rightarrow3e$ and $\mu\rightarrow e\gamma$.
  The constraints on other free parameters, such as  the mass $M_{3}$ of the third $T$-odd lepton
generation, the mixing matrix elements $(V_{Hl})_{ij}$, etc. are
very weak. So, it is possible that the LHT model can give sizable
cross sections for the
 processes $e^{+}e^{-}\rightarrow l_{i}\bar{l}_{j}$. Our calculation
 has shown that it is indeed this case.

The one-loop calculation can be carried out by summing all of these
one-loop diagrams and the results will be finite and gauge
invariant. Each loop diagram is composed of some scalar loop
functions, which are calculated by using $LoopTools$ [17]. In the
following sections, we will use the $'t$ $Hooft$-Feynman gauge to
calculate their production cross sections. Because the calculation
of the loop diagrams is too tedious and the analytical expressions
are lengthy, we will not present them here.

\begin{figure}[htb]
\begin{center}
\epsfig{file=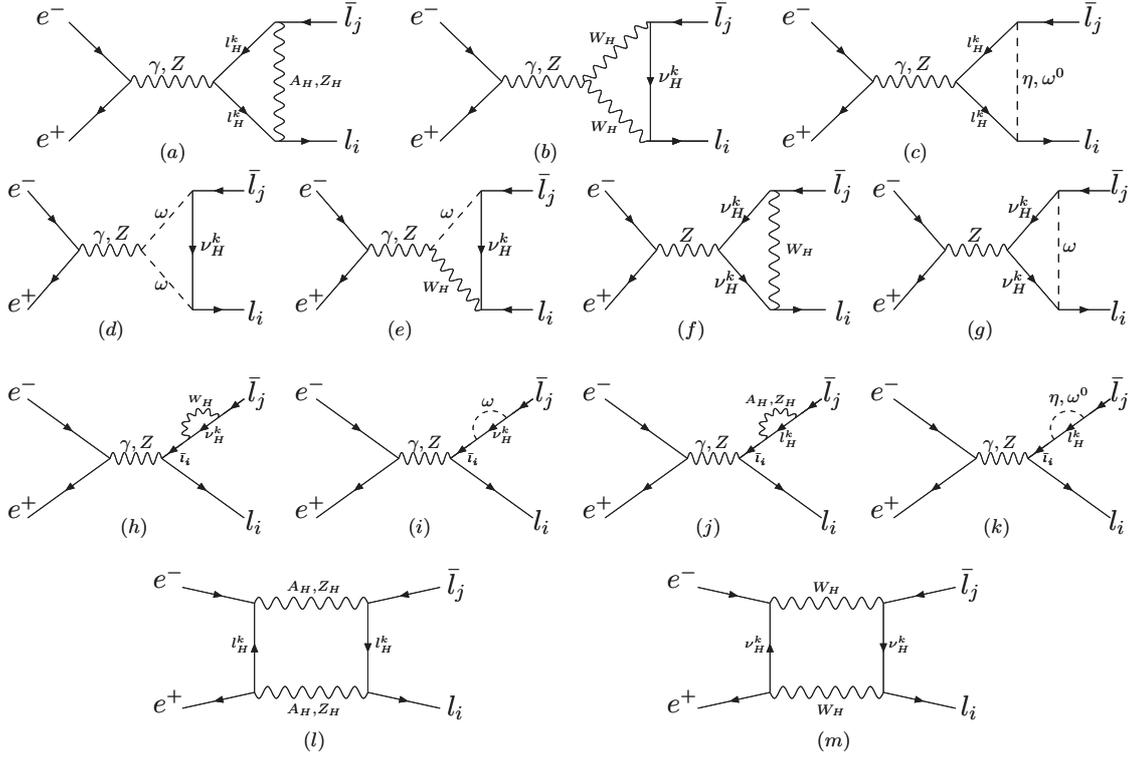,width=460pt,height=630pt} \vspace{-15cm}
\vspace{2.5cm} \caption{ The Feynman diagrams for the LFV processes
$e^{+}e^{-}\rightarrow l_{i}\bar{l}_{j}$ in the LHT model.}
\end{center}
\vspace{-0.2cm}
\end{figure}

It is obvious that, except for the SM input parameters $\alpha=
1/128.8$, $S_{W}^{2} = 0.2315$, and $M_{Z} = 91.187GeV$ [18], the
production cross sections $\sigma_{ij}$ for the LFV processes
$e^{+}e^{-}\rightarrow l_{i}\bar{l}_{j}$  are dependent on the model
dependent parameters $(V_{Hl})_{ij}$, $f$, and the $T$-odd leptons'
masses. The matrix elements $(V_{Hl})_{ij}$ can be determined
through $V_{Hl} = V_{H\nu}V_{PMNS}$. To avoid any additional
parameters introduced and to simplify our calculations, we take
$V_{Hl} = V_{PMNS}$, which means that $V_{H\nu} = I$ and the $T$-odd
leptons have no effects on the flavor violating observable in the
neutrino sector [14, 19]. Certainly, this is a very limited
scenario. However, in order to satisfy the constraints from
$\mu\rightarrow3e$ and $\mu\rightarrow e\gamma$, the mixing matrix
$V_{Hl}$ must be rather hierarchical or the first and second $T$-odd
lepton masses are quasidegenerate. Therefor, in this paper, we take
$M_{l_{H}^{1}}=M_{l_{H}^{2}}$ and $V_{Hl} = V_{PMNS}$ might be
suitable. For the PMNS matrix $V_{PMNS}$, we take the standard
parametrization form with parameters given by the neutrino
experiments [20]. Since there is not constraints on the PMNS phases,
we will take the Dirac phase to be equal to the CKM phase and set
the two Majorana phases to zero in our numerical estimations.

\begin{figure}[htb]
\begin{center}\vspace{-0.5cm}
\epsfig{file=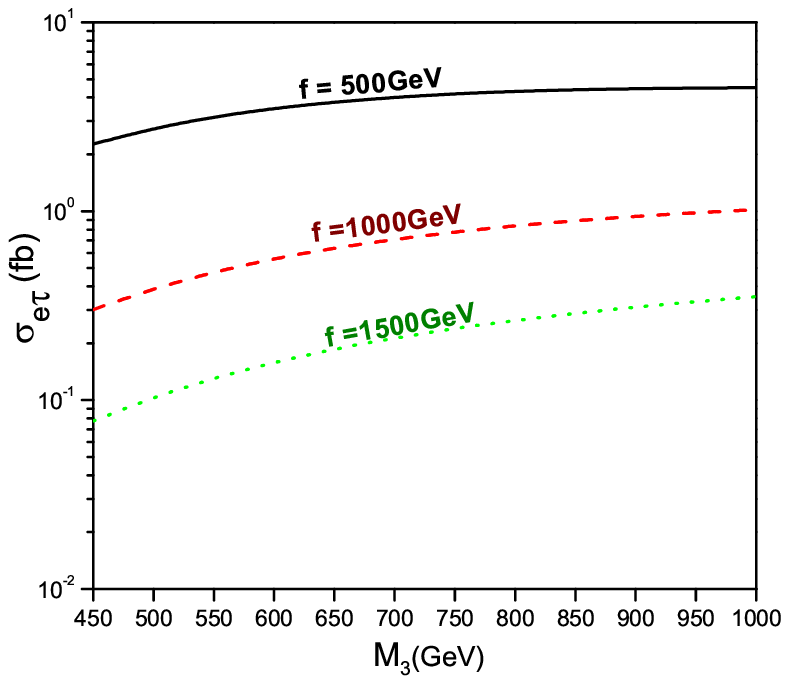,width=200pt,height=185pt}\put(-110,-5){
\textbf{\small{(a)}}}
\epsfig{file=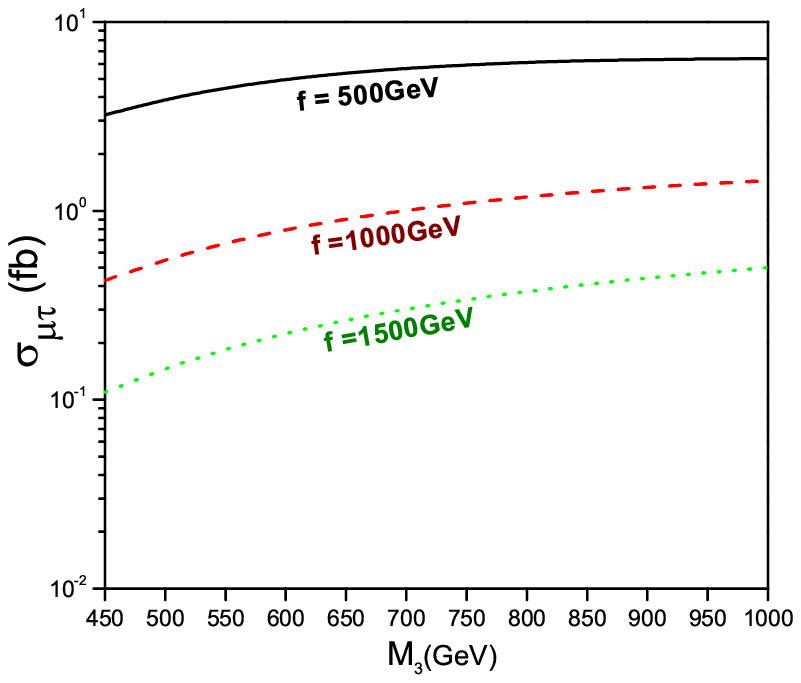,width=200pt,height=185pt} \put(-102,-5){
\small{\textbf{(b)}}} \vspace{-0.7cm}
\leftline{\hspace{0.9cm}\epsfig{file=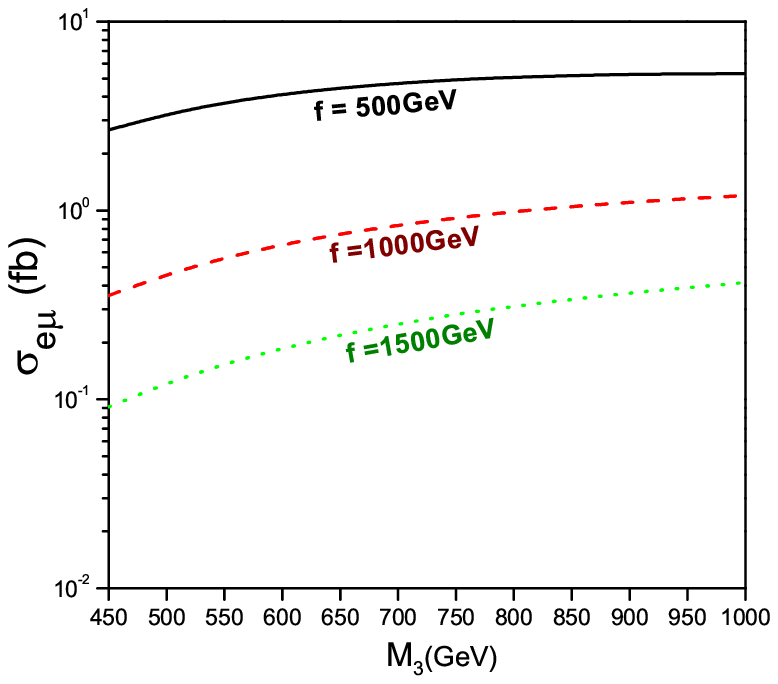,width=200pt,height=185pt}}\put(-335,-5){
\small{\textbf{(c)}}} \vspace{-0.2cm}\caption{The production cross
sections $\sigma_{ij}$ for the processes $e^{+}e^{-}\rightarrow
l_{i}\bar{l_{j}}$ as function \hspace*{2.0cm}of the mass parameter
$M_{3}$ for $M_{2}=400GeV$ and three values of the scale  \hspace*{2.0cm}
parameter $f$}
\end{center}\vspace{-0.5cm}
\end{figure}

Our numerical results are summarized in Fig.2, in which we plot the
production cross sections $\sigma _{ij}$ as functions of the mass
parameter $M_{3}$ for $M_{2}=400GeV$ and three values of the scale
parameter $f$. Figures(a), Fig.2(b) and Fig.2(c) are corresponding
to the LFV processes $e^{+}e^{-}\rightarrow e\bar{\tau}$,
$e^{+}e^{-}\rightarrow \mu\bar{\tau}$ and $e^{+}e^{-}\rightarrow
e\bar{\mu}$, respectively. In these figures, we have taken the c. m.
energy $\sqrt{s}=500GeV$. One can see from Fig.2 that the
contributions of the LHT model to the LFV processes
$e^{+}e^{-}\rightarrow l_{i}\bar{l}_{j}$ ($i\neq j$) increase as the
third family $T$-odd lepton mass $M_{3}$ increases and the scale
parameter $f$ decreases. In most of the parameter space, the
production cross sections for the states $e\bar{\tau}$,
$\mu\bar{\tau}$ and $e\bar{\mu}$ are approximately at the same order
of magnitude. For $\sqrt{s}=500GeV$, $M_{2}=400GeV$, $450GeV\leq
M_{3}\leq1000GeV$ and $500GeV\leq f \leq1500GeV$,
 there are $0.08fb\leq\sigma_{
e\bar{\tau}}\leq4.5fb$, $0.1fb\leq\sigma_{ \mu\bar{\tau}}\leq6.4fb$
and $0.09fb\leq\sigma_{e\bar{\mu}}\leq5.3fb$, respectively. If we
assume the integrated luminosity $\pounds_{int}=500fb^{-1}$, there
will be several tens and up to thousands of $l_{i}\bar{l}_{j}$
events to be generated in the future ILC experiments.

\begin{figure}[htb]
\begin{center}\vspace{-0.5cm}
\epsfig{file=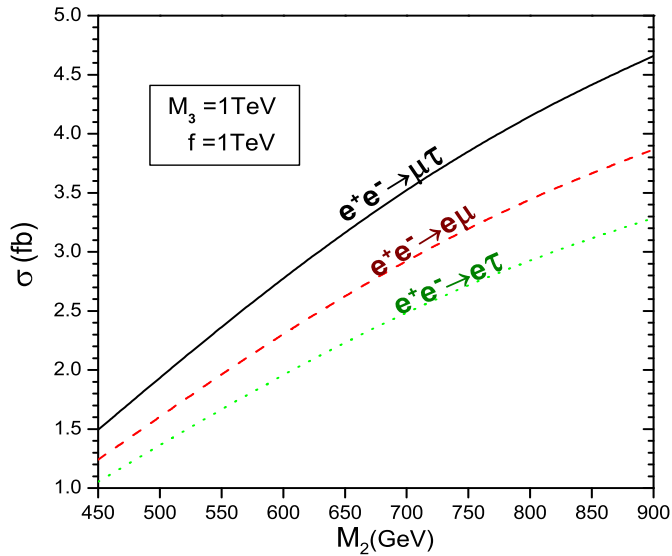,width=300pt,height=260pt} \vspace{-0.9cm}
\caption{The production cross sections $\sigma_{ij}$ for the
processes $e^{+}e^{-}\rightarrow l_{i}\bar{l_{j}}$ as function
\hspace*{2.0cm}of the mass parameter $M_{2}$ for $f=1TeV$ and
$M_{3}=1000GeV$.}
\end{center}\vspace{-0.5cm}
\end{figure}

To see the effect of the first and second generation $T$-odd lepton
masses on the production cross sections $\sigma_{ij}$, in Fig.3, we
plot  $\sigma_{ij}$ as functions of the mass parameter $M_{2}$ for
$f=1TeV$ and $M_{3}=1000GeV$. From Fig.3 one can see that the
contributions of the LHT model to the LFV processes
$e^{+}e^{-}\rightarrow l_{i}\bar{l}_{j}$  also increase as the mass
parameter $M_{2}$. For $M_{3}=1000GeV$, $f=1TeV$, $450GeV\leq
M_{2}\leq 900GeV$ and $\sqrt{s}=500GeV$, they are $1.1fb\leq\sigma_{
e\bar{\tau}}\leq3.3fb$, $1.5fb\leq\sigma_{ \mu\bar{\tau}}\leq4.7fb$
and $1.2fb\leq\sigma_{e\bar{\mu}}\leq 3.9fb$, respectively.

It is well known that, at the ILC,  the LFV production processes can
provide extremely clear signatures and are experimentally
interesting . For the three LFV processes $e^{+}e^{-}\rightarrow
e\bar{\tau}$, $e^{+}e^{-}\rightarrow \mu\bar{\tau}$ and
$e^{+}e^{-}\rightarrow e\bar{\mu}$, the final leptons always emerge
back to back and carrying a constant energy which is one-half of the
c.m. energy $\sqrt{s}$. The last process is the best one and almost
free of the SM backgrounds. For the first and second production
processes, we can assume the lepton tau decay
$\bar{\tau}\rightarrow\bar{\mu}\nu_{\mu}\bar{\nu_{\tau}}$ and
$\bar{\tau}\rightarrow\bar{e}\nu_{e}\bar{\nu_{\tau}}$, which give
rise to the signal events with opposite-sign and different-flavor
leptons and missing energy ($e\bar{\mu}+ \not\!\!E$ and
$\mu\bar{e}+\not\!\!E$). Although this kind of LFV signal is quite
spectacular, it is not free of the SM backgrounds. For example,
 the leading backgrounds of the signal $e\bar{\mu}+\not\!\!E$ mainly
come from the SM processes $e^{+}e^{-}\rightarrow
e\bar{\mu}\nu_{\mu}\bar{\nu_{e}}$ and
$e^{+}e^{-}\rightarrow\tau^{+}\tau^{-} \rightarrow
e\bar{\mu}\nu_{\mu}\bar{\nu_{e}}\nu_{\tau}\bar{\nu_{\tau}}$, which
have been discussed in Ref.[8]. They have shown that, with suitable
 cuts, the SM backgrounds can be largely suppressed. Thus, it is
possible to observe the LFV signal of the LHT model in the future
ILC experiments. Certainly, detailed study of the relevant
backgrounds is needed, which is beyond the scope of this paper.

\section*{IV. The LFV processes $\gamma\gamma\rightarrow l_{i}\bar{l}_{j}$ in the LHT model }

\hspace{0.5cm} It is well known that the ILC could offer the
possibility of working in the $\gamma\gamma$ or $e\gamma$ collision
thus realizing a very high energy photon collider [5]. For some
production processes, their cross sections at the $\gamma\gamma$
collider might be larger than the corresponding $e^{+}e^{-}$ ones;
this collider will reveal crucial information about these production
processes. To see whether the LHT model can give significant
contributions to the LFV final states $l_{i}\bar{l_{j}}$ ($i\neq j$)
via $\gamma\gamma$  collision, we will consider the LFV processes
$\gamma\gamma\rightarrow l_{i}\bar{l_{j}}$ in this section.

From the discussions given in Sec.II, we can see that the LHT model
can only induce the LFV processes $\gamma\gamma\rightarrow
l_{i}\bar{l_{j}}$ at loop level. The relevant Feynman diagrams are
shown in Fig.4. The Feynman diagrams created by exchanging the
initial photons, which are not shown in Fig.4, are also involved in
our calculations.

We use $\theta$ to denote the scattering angle between one of the
photons and one of the final leptons. Then, in the center of mass
(c.m.) system, we express all the four-momenta of the initial and
final particles by means of the $\gamma\gamma$ c.m. energy
$\sqrt{\hat{s}}$ and the scattering angle $\theta$. The
four-momentum components ($E$, $p_{x}$, $p_{y}$, $p_{z}$) of final
particles $l_{i}$ and $\bar{l}_{j}$ can be written as
\begin{eqnarray}
&&p_{1}=(E_{l_{i}}, \sqrt{E_{l_{i}}^{2}-m^{2}_{l_{i}}}\sin\theta, 0,
\sqrt{E_{l_{i}}^{2}-m^{2}_{l_{i}}}\cos\theta ),\\
&&p_{2}=(E_{\bar{l}_{j}},-
\sqrt{E_{\bar{l}_{j}}^{2}-m^{2}_{\bar{l}_{j}}}\sin\theta, 0,-
\sqrt{E_{\bar{l}_{j}}^{2}-m^{2}_{\bar{l}_{j}}}\cos\theta ),
\end{eqnarray}
\hspace{2.7cm}

where $E_{l_{i}}=E_{\bar{l}_{j}}=\sqrt{\hat{s}}/2$. The Mandelstam
variables are defined as\vspace{-0.1cm}
\begin{eqnarray}
\nonumber
\hat{s}=(k_{1}+k_{2})^{2}=(p_{1}+p_{2})^{2},\\
\nonumber
\hat{t}=(k_{1}-p_{1})^{2}=(k_{2}-p_{2})^{2},\\
\nonumber
 \hat{u}=(k_{1}-p_{2})^{2}=(k_{2}-p_{1})^{2}.\\
 \nonumber
\end{eqnarray}

\begin{figure}[htb]
\begin{center}
\vspace{-2.0cm} \epsfig{file=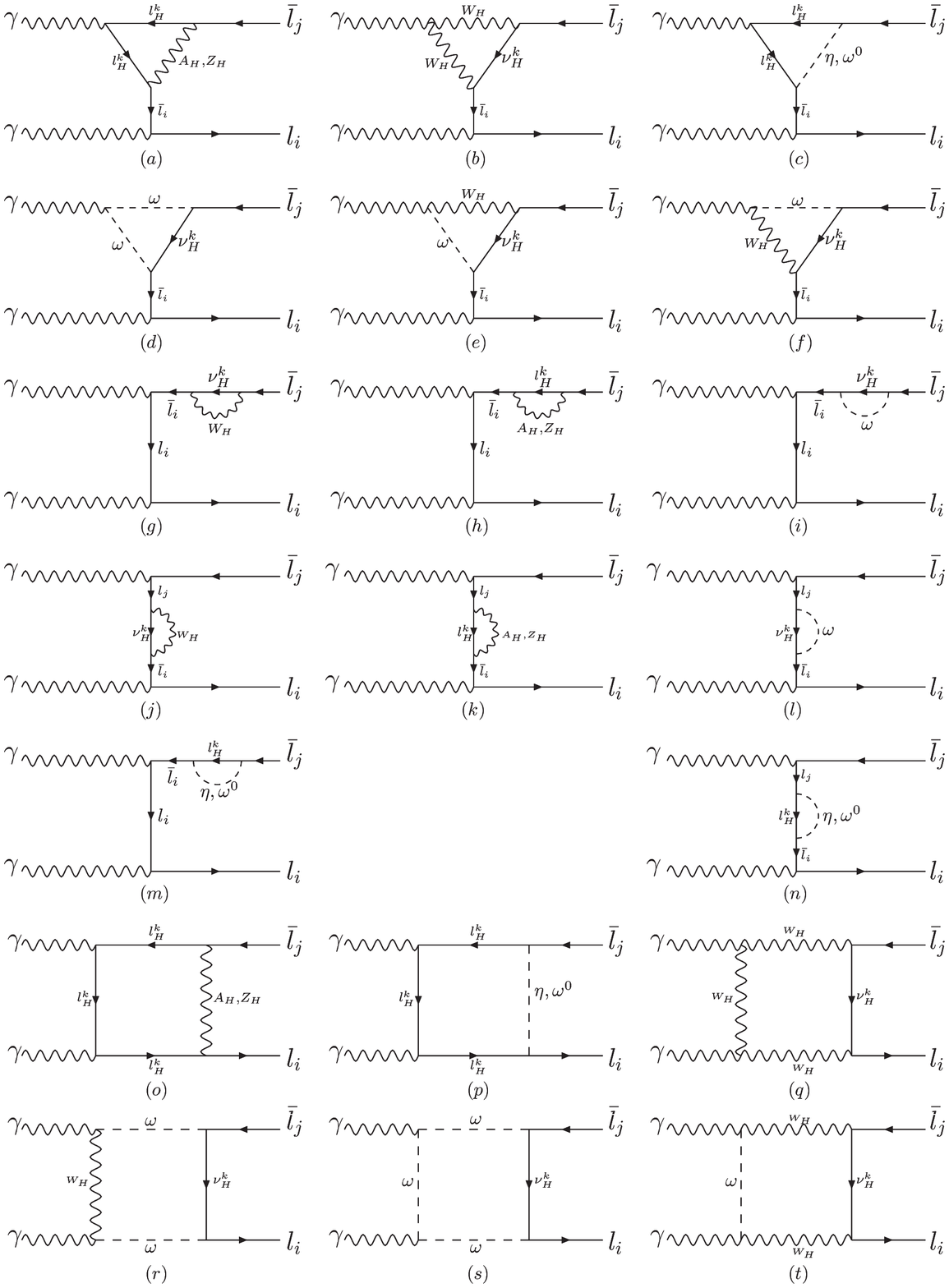,width=420pt,height=550pt}
\vspace{-2.3cm} \caption{ The Feynman diagrams for the LFV processes
$\gamma\gamma\rightarrow l_{i}\bar{l}_{j}$ in the LHT model.
\hspace*{1.8cm}The diagrams obtained by exchanging the initial
photons are not shown here.}
\end{center}
\vspace{-0.5cm}
\end{figure}

Where $k_{1}$ and $k_{2}$ are the 4-momenta of the initial photons,
which can be written as $k_{1}=(\sqrt{\hat{s}}/2, 0, 0,
\sqrt{\hat{s}}/2)$ and $k_{2}=(\sqrt{\hat{s}}/2, 0, 0,
-\sqrt{\hat{s}}/2)$. With the above definitions, we can give the
renormalized amplitudes of the LFV processes
$\gamma\gamma\rightarrow l_{i}l_{j}$. To simplify our paper, we do
not give their explicit expressions here. The cross section for the
LFV process $\gamma\gamma\rightarrow l_{i}l_{j}$ can be generally
expressed in the form
\begin{eqnarray} \hat{\sigma}(\hat{s})=\frac{1}{16\pi
\hat{s}^{2}}
\int^{\hat{t}^{+}}_{\hat{t}^{-}}d\hat{t}\overline{\sum_{spin}}|\mathcal
{M}|^{2},
\end{eqnarray}
where
$\hat{t}^{\pm}=\frac{1}{2}[(m_{l_{i}}^{2}+m_{\bar{l}_{j}}^{2}-\hat{s})\pm\sqrt{(m_{l_{i}}^{2}+m_{\bar{l}_{j}}^{2}-\hat{s})^{2}
-4m_{l_{i}}^{2}m_{\bar{l}_{j}}^{2}}]$, and the bar over summation
means to take the average over the initial polarizations of the
photons.

At the ILC, the effective cross section of the subprocess
$\gamma\gamma\rightarrow l_{i}\bar{l}_{j}$ can be written as
\begin{eqnarray}
\sigma(s)=\int^{x_{max}}_{E_{0}/\sqrt{s}}dz\frac{d\mathcal
{L}_{\gamma\gamma}}{dz}\hat{\sigma}_{\gamma\gamma\rightarrow l_{i}\bar{l}_{j}}(\hat{s}=z^{2}s),
\end{eqnarray}
where $E_{0}=m_{l_{i}}+m_{\bar{l}_{j}}$, and
$\sqrt{s}(\sqrt{\hat{s}})$ is the $e^{+}e^{-}(\gamma\gamma)$ c.m.
energy; $d\mathcal {L}_{\gamma\gamma}/{dz}$ is the photon-beam
luminosity distribution, which is defined as
\begin{eqnarray}
\frac{d\mathcal
{L}_{\gamma\gamma}}{dz}=2z\int^{x_{max}}_{z^{2}/x_{max}}\frac{dx}{x}F_{\gamma/e}(x)F_{\gamma/e}(z^{2}/x).
\end{eqnarray}
For the initial unpolarized electron and laser-photon beams, the
energy spectrum of the backscattered photon is given by [21]
\begin{eqnarray}
F_{\gamma/e}(x)=\frac{1}{D(\xi)}[1-x+\frac{1}{1-x}-\frac{4x}{\xi(1-x)}+\frac{4x^{2}}{\xi^{2}(1-x)^{2}}]
\end{eqnarray}
with
\begin{eqnarray}
D(\xi)=(1-\frac{4}{\xi}-\frac{8}{\xi^{2}})\ln(1+\xi)+\frac{1}{2}+\frac{8}{\xi}-\frac{1}{2(1+\xi)^{2}},
\end{eqnarray}
where $\xi=4E_{0}\omega_{0}/{m_{e}^{2}}$, $m_{e}$ and $E_{0}$ are
the incident electron mass and energy, respectively. $\omega_{0}$ is
the laser-photon energy, $x$ is the fraction of the energy of the
incident electron carried by the backscattered photon. In order to
spoil the creation of $e^{+}e^{-}$ pair by the interaction of the
incident and backscattered photons, in our calculation, we require
$\omega_{0}x_{max}\leq m_{e}^{2}/E_{e}$, which implies $\xi\leq4.8$.
For the choice $\xi=4.8$, there are $x_{max}\simeq0.83$ and
$D(\xi)=1.8$.

\vspace{0.8cm}
\begin{figure}[htb]
\begin{center}\vspace{-1cm}
\epsfig{file=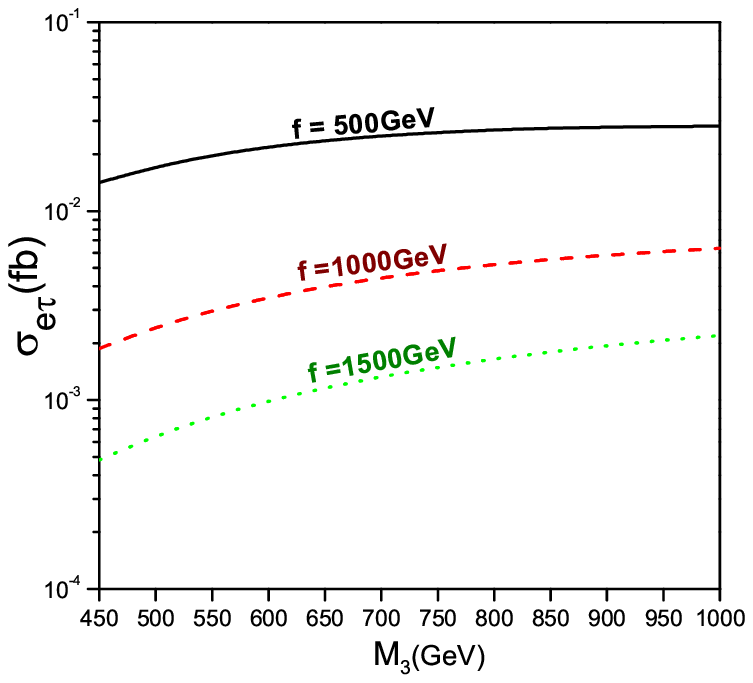,width=200pt,height=185pt}\put(-110,-5){
\textbf{\small{(a)}}}
\epsfig{file=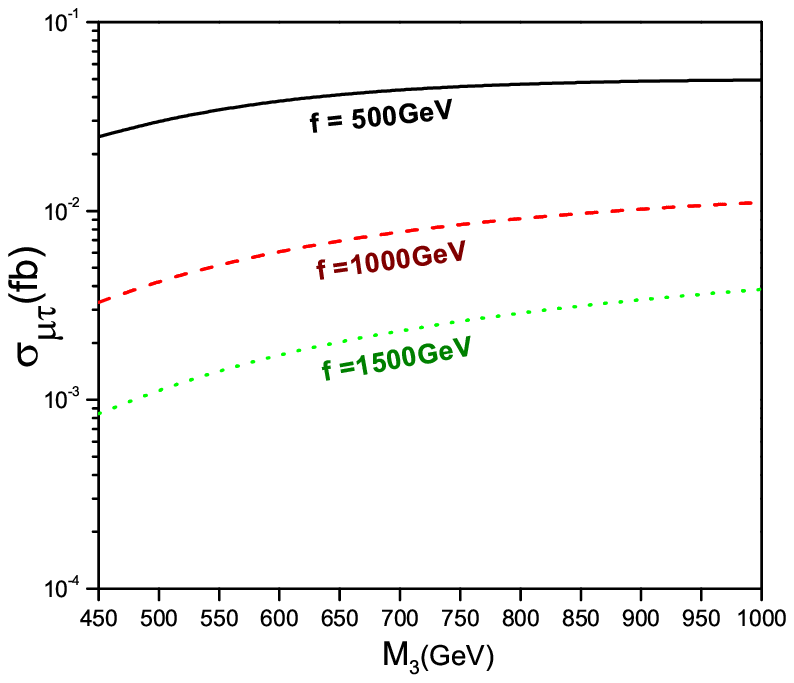,width=200pt,height=185pt} \put(-102,-5){
\small{\textbf{(b)}}} \vspace{-0.7cm}
\leftline{\hspace{0.9cm}\epsfig{file=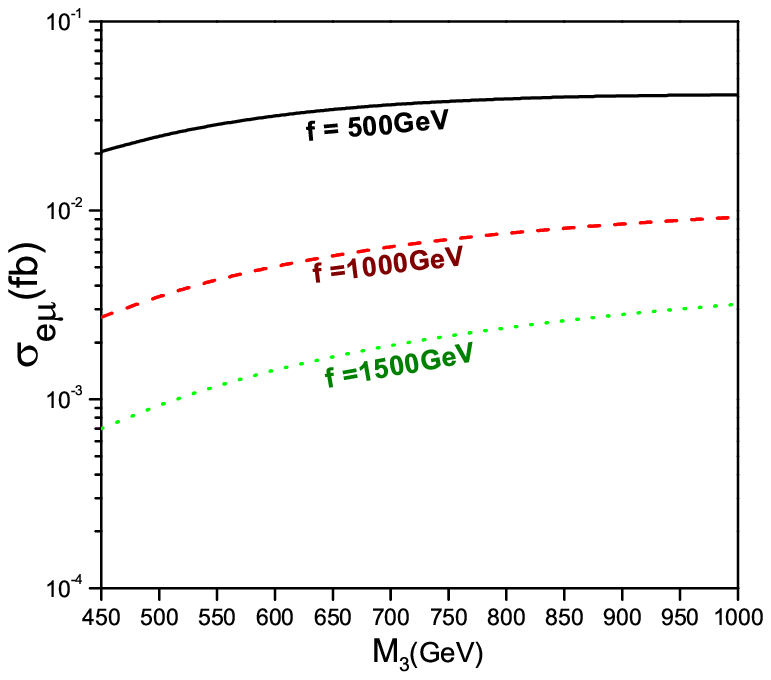,width=200pt,height=185pt}}\put(-335,-5){
\small{\textbf{(c)}}} \vspace{-0.2cm}\caption{The production cross
sections $\sigma_{ij}$ for the subprocesses $\gamma\gamma\rightarrow
l_{i}\bar{l_{j}}$ as functions \hspace*{2.0cm}of the mass parameter
$M_{3}$ for $M_{2}=400GeV$ and three values of the scale\newline \hspace*{2.0cm}parameter $f$.}
\end{center}
\end{figure}
Similarly with Sec.III, we also assume $V_{H\nu}=I$ and that the
masses of the first generation $T$-odd fermions are equal to those
of the second generation $T$-odd fermions. In this case, our
numerical results are summarized in Figs.5 and Fig.6. Figure 5 plots
the effective production cross sections $\sigma_{ij}$ as functions
of the mass parameter $M_{3}$ for $M_{2}=400GeV$ and three values of
the scale parameter $f$, while Fig.6 plots the cross sections
$\sigma_{ij}$ as functions of $M_{2}$ for $f=500GeV$ and
$M_{3}=1000GeV$. One can see from these figures that, in most of the
parameter space, the contributions of the LHT model to the LFV
processes $e^{+}e^{-}\rightarrow \gamma\gamma\rightarrow
l_{i}\bar{l_{j}}$ are smaller than those for the LFV processes
$e^{+}e^{-}\rightarrow l_{i}\bar{l_{j}}$. With the increasing of the
mass parameter $M_{i}$ and the decreasing of the scale parameter
$f$, the effective cross sections of the LFV processes
$\gamma\gamma\rightarrow l_{i}\bar{l_{j}}$ become larger. Such a
behavior is similar to that
 for the LFV processes $e^{+}e^{-}\rightarrow
l_{i}\bar{l_{j}}$. For $\sqrt{s}=500GeV$, $M_{2}=400GeV$,
$450GeV\leq M_{3}\leq1000GeV$ and $500GeV\leq f \leq1500GeV$, there
are $6\times 10^{-4}fb\leq\sigma_{ e\bar{\tau}}\leq3\times
10^{-2}fb$, $8\times 10^{-4}fb\leq\sigma_{ \mu\bar{\tau}}\leq5\times
10^{-2}fb$ and $7\times 10^{-4}fb\leq\sigma_{ e\bar{\mu}}\leq4\times
10^{-2}fb$, respectively. There will be several tens of
$l_{i}\bar{l}_{j}$ events to be generated in the future ILC
experiment with $\pounds_{int}=500fb^{-1}$ and $\sqrt{s}=500GeV$.

\begin{figure}[htb]
\begin{center}\vspace{-0.5cm}
\epsfig{file=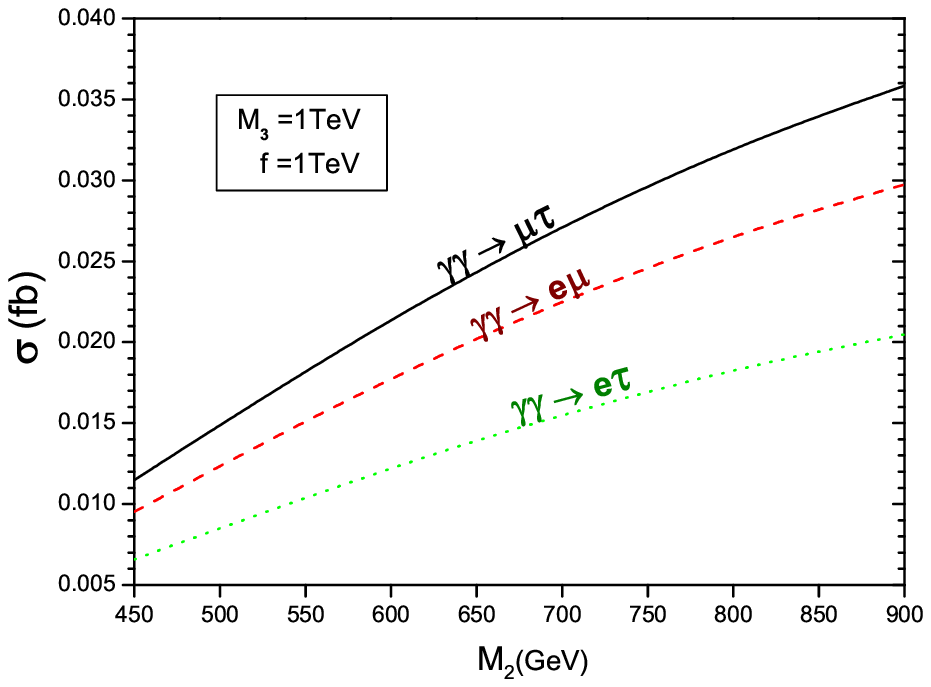,width=300pt,height=260pt} \vspace{-0.9cm}
\caption{The production cross sections $\sigma_{ij}$ for the
subprocesses $\gamma\gamma\rightarrow l_{i}\bar{l_{j}}$ as functions
\hspace*{2.0cm}of the mass parameter $M_{2}$ for $f=1TeV$ and
$M_{3}=1000GeV$.}
\end{center}\vspace{-0.5cm}
\end{figure}

Similar with the LFV process $e^{+}e^{-}\rightarrow e\bar{\mu}$, the
LFV process $\gamma\gamma\rightarrow e\bar{\mu}$ is almost free of
the SM background. For the LFV processes $\gamma\gamma\rightarrow
e\bar{\tau}$ and $\gamma\gamma\rightarrow\mu\bar{\tau}$, the SM
backgrounds mainly come from the processes $\gamma\gamma\rightarrow
\tau^{+}\tau^{-}$ and $\gamma\gamma\rightarrow WW$ with the lepton
$\tau$ and the electroweak gauge boson $W$ leptonic decaying. It has
been shown that, by applying appropriate kinematical cuts, the
backgrounds can be significantly suppressed and the ratio of signal
to background would be enhanced [6]. Thus, the LFV signatures of the
LHT model might have a chance of being observed via the subprocesses
$\gamma\gamma\rightarrow l_{i}\bar{l_{j}}$ in the future ILC
experiments.

\section*{V. Conclusions and discussions}

\hspace{0.5cm}The LHT model is one of the attractive little Higgs
models. To simultaneously implement $T$-parity, the LHT model
introduces new mirror fermions ($T$-odd quarks and $T$-odd leptons).
The flavor mixing in the mirror fermion sector gives rise to a new
source of flavor violation, which might generate significant
contributions to some flavor violation processes.

The evidence for the neutrino masses and flavor mixing, which can be
seen as the first experimental clue of new physics beyond the SM,
implies the nonconservation of the lepton flavor symmetry. Thus, the
LFV processes related charged leptons are expected, which are very
sensitive to new physics beyond the SM. Taking into account the
constraints on the free parameters of the LHT model from the rare
decay processes $\mu\rightarrow e\gamma$ and $\mu\rightarrow 3e$, in
this paper, we have considered the contributions of the LHT model to
the LFV processes $e^{+}e^{-}\rightarrow l_{i}\bar{l}_{j}$ and
$\gamma\gamma\rightarrow l_{i}\bar{l}_{j}$ ($i\neq j$), which are
induced at the one-loop level and will be of interest for the future
ILC experiments. We find that, in wide range  of the parameter
space, the production cross section of the LFV process
$e^{+}e^{-}\rightarrow l_{i}\bar{l}_{j}$ can reach several $fb$, and
that of the LFV process $\gamma\gamma\rightarrow l_{i}\bar{l}_{j}$
can reach the order of magnitude of $10^{-2}fb$. This means that
there will be several and up to thousands of $l_{i}\bar{l}_{j}$
events to be generated each year for the designed luminosity of
$\pounds_{int}=500fb^{-1}$ at the ILC. Since the
 production rate of these LFV processes predicted by the SM is almost negligible, the
observation of such $l_{i}\bar{l}_{j}$ events would be a robust
evidence of the LHT model. Therefore, these LFV processes may serve
as a sensitive probe of the LHT model.

An important tool of the ILC is the use of the polarized beams. One
expects that a high polarization degree between $80\%$ and $90\%$
can be reached [3,4]. Beam polarization is not only useful for a
possible reduction of the background, but might also serve as a
possible tool to disentangle different contributions to the signal
and to directly analyze the interaction structure of new physics
models. In our calculation, we have not considered the polarization
of the incident electron. Certainly, if we consider this case, our
numerical results will be changed. Furthermore,  we can also assume
$V_{Hl}=V_{CKM}$, which makes the values of the production cross
sections of the LFV processes $e^{+}e^{-}\rightarrow
l_{i}\bar{l}_{j}$ different from those for assuming $V_{Hl} =
V_{PMNS}$. However, our physical conclusions are not changed.

It is obvious that the production cross sections of the LFV
processes $e^{+}e^{-}\rightarrow l_{i}\bar{l}_{j}$ and
$\gamma\gamma\rightarrow l_{i}\bar{l}_{j}$ are dependent on the
factor $\Sigma_{\alpha} V^{\alpha i \ast}_{Hl}  V^{\alpha j}_{Hl}$
with $\alpha=1, 2, 3$. For assuming $V_{Hl} = V_{PMNS}$, the PMNS
phases should have effects on these cross sections. In our numerical
estimations, we have fixed the values of the PMNS phases. If we vary
these values, the numerical results for the cross sections $\sigma_{
e\bar{\tau}}$, $\sigma_{ \mu\bar{\tau}}$ and $\sigma_{e\bar{\mu}}$
are also changed. However, these variations are not significantly
large. Our physical conclusions are not changed.

The $T$-odd quarks predicted by the LHT model also have new flavor
violating interactions with the SM fermions mediated by the new
gauge bosons and at higher order by the triplet scalar. These
interactions are governed by the new mixing matrix $V_{Hd}$ for
down-quarks. So the $T$-odd quark sector can also generate
significant contributions to some flavor changing processes [12, 13,
22, 23]. For example, Ref.[24] has shown that it is possible to test
the signatures of the LHT model at the ILC and LHC experiments via
the flavor changing processes $e^{+}e^{-}\rightarrow \bar{t}c$,
$\gamma\gamma\rightarrow \bar{t}c$ and $pp\rightarrow \bar{t}c$.

The $T$-odd particles predicted by the LHT model can only be
produced in pairs and give the direct signals at the LHC, which have
close resemblance with those of supersymmetry with conserved $R$
parity or universal extra dimensions with $KK$ parity. The
possibility of observing the $T$-odd leptons  at the LHC has been
studied in Ref.[25]. If the $T$-odd leptons are found at the LHC, we
expect that our work will be helpful to determine their masses and
couplings with high accuracy at the ILC.

\section*{Acknowledgments} \hspace{5mm}This work was
supported in part by the National Natural Science Foundation of
China under Grants No.10975067, the Specialized Research Fund for
the Doctoral Program of Higher Education (SRFDP) (No.200801650002).


\vspace{1.0cm}


\begin{thebibliography}{99}

\bibitem{y1}C. K. Jung {\em et al.}, {\em Annu. Rev. Nucl. Part. Sci.} {\bf 51}, 451 (2001);
            Q. R. Ahmad {\em et al.} [SNO Collaboration], {\em Phys. Rev. Lett.}
            {\bf 89}, 011301 (2002); K. Eguchi {\em et al.} [ KamLAND
            Collaboration], {\em Phys. Rev. Lett.} {\bf 90}, 021802 (2003); M.
            H. Ahn {\em et al.} [K2K Collaboration], {\em Phys. Rev. Lett.} {\bf
            90}, 041801 (2003).

\bibitem{y2}For recent reviews on neutrino physics, see {\em e.g.}, V. Barger, D.Marfatia, and K.Whisnant,
            {\em Int. J. Mod. Phys. E} {\bf12}, 569 (2003); R. N. Mohapatra and
            A. Y. Smirnov, {\em Annu. Rev. Nucl. Part. Sci.} {\bf56}, 569
            (2006); A. Strumia and F. Vissani, {\em hep-ph}/{\bf0606054}; M. C.
            Gonzalez-Garcia and M. Maltoni, {\em Phys. Rep.} {\bf460}, 1 (2008);
            Z. Z. Xing, {\em Int. J. Mod. Phys. A} {\bf23}, 4255 (2008).

\bibitem{y3}T. Abe {\em et al.} [American Linear Collider Group], {\em hep-ex}/{\bf0106057};
            J. A. Aguilar-Saavedra {\em et al.} [ECFA/DESY LC Physics Working
            Group ], {\em hep-ph}/{\bf0106315}; Koh Abe {\em et al.} [ACFA
            Linear Collider Working Group ], {\em hep-ph}/{\bf0109166}; ILC
            Technical Review Committee, second report, 2003, SLAC-R-606,
            February 2003.

\bibitem{y3}G. Aarons {\em et al.} [ILC Collaboration],
            "International Linear Collider Reference Design Report Volume 2:
            Physics at the ILC," \emph{arXiv:} \textbf{0709.1893}; Brau et al.
            [ILC Collaboration], "ILC Reference Design Report Volume 1 -
            Executive Summary," \emph{arXiv:} \textbf{0712.1950}.

\bibitem{y5}I. F. Ginzburg, \emph{arXiv:} \textbf{0912.4841}.

\bibitem{y6}M. Cannoni, S. Kolb and O. Panella, {\em Phys. Rev. D} {\bf68}, 096002 (2003);
            F. Deppisch {\em et al.}, {\em Phys. Rev. D} {\bf69}, 054014 (2004);
            Y. B. Sun {\em et al.}, {\em JHEP} {\bf0409}, 043 (2004); M.
            Cannoni, C. Carimalo, W. Da Silva and O. Panella, {\em Phys. Rev. D}
            {\bf72}, 115004 (2005); {\em Erratum-ibid., D} {\bf72}, 119907
            (2005); M. Cannoni and O. Panella, {\em Phys. Rev. D} {\bf79},
            056001 (2009); J. Cao, L. Wu, J. Yang, {\em Nucl. Phys. B} {\bf829},
            370 (2010).


\bibitem{y7}Chong-Xing Yue, Yan-Ming Zhang and Hong Li,  {\em J. Phys. G} {\bf29}, 737 (2003);
            Guo-Li Liu, \emph{arXiv:} \textbf{1002.0659}.

\bibitem{y8}P. M. Ferreira, R. B. Guedes, and R. Santos, {\em Phys. Rev. D} {\bf75}, 055015 (2007);
            J. I. Aranda  {\em et al.}, {\em Phys. Rev. D} {\bf79}, 093009
            (2009).

\bibitem{y9} H. C. Cheng, I. Low, {\em JHEP} {\bf0309}, 051 (2003); {\em JHEP} {\bf0408}, 061 (2004); I. Low,
             {\em JHEP} {\bf0410}, 067 (2004).

\bibitem{y10} M. Schmaltz and D. Tucker-Smith, {\em Ann. Rev. Nucl. Part. Sci.} {\bf55}, 229 (2005);
              M. Perelstein, {\em Prog. Part. Nucl. Phys.} {\bf58}, 247 (2007).

\bibitem{y11} J. Hubisz and P. Meade, {\em Phys. Rev. D} {\bf71}, 035016 (2005);
              A. Belyaev, Chuan-Ren Chen, K. Tobe, C. P. Yuan, {\em Phys. Rev. D}
              {\bf74}, 115020 (2006); J. Hubisz, P. Meade, A. Noble and M.
              Perelstein, {\em JHEP} {\bf0601}, 135 (2006); J. Hubisz, S. J. Lee
              and G. Paz, {\em JHEP} {\bf0606}, 041 (2006); M. Blanke, {\em et
              al.}, {\em JHEP} {\bf0612}, 003 (2006).

\bibitem{y12}M. Blanke {\em et al.}, {\em JHEP} {\bf0701}, 066 (2007).

\bibitem{y13}T. Goto, Y. Okada and Y. Yamamoto, {\em Phys. Lett. B} {\bf670}, 378 (2009).

\bibitem{y14}F. del Aguila, J. I. Illana and M. D. Jenkins, {\em JHEP} {\bf0901}, 080 (2009);
             M. Blanke {\em et al.}, {\em Acta Phys. Polon. B} {\bf41}, 657 (2010).

\bibitem{y15}S. R. Choudhury, A. S. Cornell, A. Deandrea, N. Gaur and A. Goyal, {\em Phys. Rev. D} {\bf75}, 055011 (2007);
             M. Blanke, A. J. Buras, B. Duling, A. Poschenrieder and C.
             Tarantino, {\em JHEP} {\bf0705}, 013 (2007).

\bibitem{y16}Chong-Xing Yue, Jin-Yan Liu, Shi-Hai Zhu, {\em Phys. Rev. D} {\bf78}, 095006 (2008);
             Wei Liu, Chong-Xing Yue, Jiao Zhang, {\em Eur. Phys. J. C} {\bf68},
             197 (2010).

\bibitem{y17}T. Hahn, M. Perez-Victoria, {\em Computl. Phys. Commun. }{\bf118}, 153 (1999);
             T. Hahn, {\em Nucl. Phys. Proc. Suppl.} {\bf135}, 333 (2004).

\bibitem{y18}W. M. Yao et al. [Particle Data Group], {\em J. Phys. G}{\bf 33}, 1(2006)
             and partial updat for the 2008 edition.

\bibitem{y19}M. Blanke, A. J. Buras, {\em Acta Phys. Polon. B} {\bf38}, 2923 (2007);
             Chong-Xing Yue, Nan Zhang, Shi-Hai Zhu, {\em Eur. Phys. J.
             C} {\bf53}, 215 (2008).

\bibitem{y20}O. Mena, S. J. Parke, {\em Phys. Rev. D} {\bf69}, 117301(2004); J. D. Bjorken, P. F. Harrison,
             W. G. Scott, {\em Phys. Rev. D} {\bf74}, 073012 (2006); R. N.
             Mohapatra {\em et al.}, {\em Rep. Prog. Phys.} {\bf70}, 1757 (2007);
             C. Giunti, {\em Nucl. Phys. B, Proc. Suppl.} {\bf169}, 309 (2007).

\bibitem{y21}F. Ginzburg {\em et al.}, {\em Nucl.
             Instrum} {\bf219}, 5 (1984); V. Telnov, {\em Nucl.
             Instrum. Methods Phys. Res. A} {\bf294}, 72 (1990).

\bibitem{y22}M. Blanke {\em et al.}, {\em Phys. Lett. B} {\bf646},
             253 (2007); M. Blanke and A. J. Buras, {\em Acta Phys. Polon.
             B} {\bf38}, 2923 (2007).

\bibitem{23}Wei Liu, Chong-Xing Yue, Hui-Di Yang, {\em Phys. Rev. D} {\bf 79},
            034008 (2009); Chong-Xing Yue, Jiao Zhang, Wei Liu, {\em Nucl. Phys. B} {\bf832},
             342(2010)[ \emph{arXiv:}
            \textbf{1002.2010}].

\bibitem{y24}X. L. Wang, Y. J. Zhang, H. L. Jin, Y. H. Xi, {\em Nucl. Phys. B} {\bf807},
             210 (2009); {\bf810}, 226 (2009); J. I. Aranda, A. Cordero-Cid,  F.
             Ramirez-Zavaleta,  J. J. Toscano, E. S. Tututi, {\em Phys. Rev. D} {\bf
             81}, 077701 (2010).

\bibitem{y25}G. Cacciapaglia, A. Deandrea, S.Rai Choudhury,
             N Gaur, {\em Phys. Rev. D} {\bf 81},
             075005 (2010).

\end{thebibliography}
\end{document}